\documentstyle[sprocl,epsf,rotating]{article}

\newcommand{\lsim}{\raisebox{-0.13cm}{~\shortstack{$<$ \\[-0.07cm] $\sim$}}~}

\newcommand{\beq}{\begin{equation}}
\newcommand{\eeq}{\end{equation}}
\newcommand{\bea}{\begin{eqnarray}}
\newcommand{\eea}{\end{eqnarray}}
\newcommand{\non}{\nonumber}
\newcommand{\squ}{\tilde{q}}
\newcommand{\gl}{\tilde{g}}

\newcommand{\sq}{\tilde{q}}
\newcommand{\sqb}{\bar{\tilde{q}}}
 \newcommand{\zp}[3]{{Z.\ Phys.} {\bf #1} (19#2) #3}
 \newcommand{\np}[3]{{Nucl.\ Phys.} {\bf #1} (19#2)~#3}
 \newcommand{\pl}[3]{{Phys.\ Lett.} {\bf #1} (19#2) #3}
 \newcommand{\pr}[3]{{Phys.\ Rev.} {\bf #1} (19#2) #3}
 \newcommand{\prl}[3]{{Phys.\ Rev. Lett.} {\bf #1} (19#2) #3}

\input{axodraw.sty}
\def\be{\begin{equation}}
\def\ee{\end{equation}}
\def\bea{\begin{eqnarray}}
\def\eea{\end{eqnarray}}

\begin{document}

\setcounter{page}{0}

\begin{large}
\begin{flushright}
CERN-TH/97-325 \\
hep-ph/9711408 \\
November 1997
\end{flushright}
\end{large}

\vspace*{1cm}

\renewcommand{\thefootnote}{\fnsymbol{footnote}}

\begin{center}
{\large \bf Squark and Gluino Production at Hadron
Colliders \footnote{Contribution
to the proceedings of the {\it International Workshop on Quantum Effects
in the MSSM}, 9--13 September 1997, Barcelona, Spain}}

\vspace*{1cm}

{\large Michael Spira}

\vspace*{1cm}

{\large \it CERN, Theory Division, CH-1211 Geneva 23, Switzerland}

\vspace*{2cm}
{\large \bf Abstract}
\end{center}

\begin{large}
\noindent
The theoretical status of squark and gluino production at present and future
hadron colliders is reviewed. I shall concentrate on the evaluation of
SUSY-QCD corrections to the production cross sections and their
phenomenological implications.

\vspace*{\fill}

\begin{flushleft}
CERN-TH/97-325 \\
hep-ph/9711408 \\
November 1997
\end{flushleft}
\end{large}

\thispagestyle{empty}

\newpage

\title{Squark and Gluino Production at Hadron Colliders}

\author{MICHAEL SPIRA}

\address{CERN, Theory Division, CH-1211 Geneva 23, Switzerland \\
E-mail: Michael.Spira@cern.ch}


\maketitle\abstracts{
The theoretical status of squark and gluino production at present and future
hadron colliders is reviewed. I shall concentrate on the evaluation of
SUSY-QCD corrections to the production cross sections and their
phenomenological implications.
}

\section{Introduction}
Supersymmetry is a new symmetry between fermions and bosons, which has not been
observed so far at the level of elementary particles. The symmetry is
motivated by the interest of providing a natural solution to the hierarchy
problem due to the
absence of quadratic divergences in perturbation theory, in contrast with the
Standard Model. Supersymmetric grand unified theories
have predicted a Weinberg angle in striking agreement with the present
measurements at the electroweak scale. An important constraint of
supersymmetric theories is that they
predict a light scalar Higgs boson with mass $M_h\lsim 130$ GeV.

The search for Higgs bosons and supersymmetric particles provides a strong
motivation for present and future experiments. The colored supersymmetric
particles, squarks and gluinos, can be searched for at the present Tevatron
experiments, a $p\bar p$ collider with a c.m.\ energy of 1.8 TeV upgraded soon
to 2 TeV, and the LHC in the near future, a $pp$ collider with a c.m.\ energy
of 14 TeV. The Tevatron searches set presently the most stringent mass bounds
on these particles: at 95\% CL,
gluinos have to be heavier than $\sim 175$ GeV, while squarks with masses below
$\sim 175$ GeV have been excluded for gluino masses $m_{\gl} \lsim 300$ GeV
\cite{bounds}.
In $R$-parity conserving theories, e.g.\ the minimal supersymmetric extension of
the Standard Model [MSSM], supersymmetric particles can only be produced in
pairs. All supersymmetric particles will decay to the lightest one
[the LSP], which is stable stable thanks to conserved $R$-parity. As a
consequence the
typical signatures for the production of supersymmetric particles will mainly
be jets and missing transverse energy, carried away by the LSP.

Squarks and gluinos can be produced in 4 different processes,
\beq
p\bar p/pp \to \squ \squ, \squ \bar{\squ}, \gl \gl, \squ \gl + X
\label{eq:sqglprod}
\eeq
We worked in the approximation of mass-degenerate light-flavored squarks, so
that mixing effects are absent. However, this approximation is not valid for
stop production, where mass splitting and mixing effects are sizeable. Stop
production has been presented elsewhere \cite{stops}. The light quarks
($u,d,s,c,b$)
will be treated as massless particles, while the top quark will be included in
loops with a mass $m_t=175$ GeV. Moreover, stops are taken into account in the
virtual
loops, too, with a mass equal to the other squark masses. Since their effect
will be suppressed, this provides a reasonable approximation. In
Eq.~\ref{eq:sqglprod} a summation over all possible squark flavors and charge
conjugate final states should be implicitly understood.

\section{Lowest Order}
The calculation of the lowest order (LO) cross sections of the processes in 
Eq.~\ref{eq:sqglprod} was performed a long time ago \cite{lo}. The
4 production processes divide into several partonic subprocesses [$N_F=5$]:
\begin{description}
\item[(i)] \underline{$qq' \to \squ \squ'$}: \\
\beq
\hat\sigma_{LO} = \frac{4\pi\alpha_s^2}{9\hat s} \left\{
\frac{\hat s + 2 m_-^2}{\hat s} L_1
- \beta_{\squ} \frac{m_{\gl}^2 \hat s + 2 m_-^4}{m_{\gl}^2 \hat s + m_-^4}
- \delta_{qq'} \frac{2}{3} \frac{m_{\gl}^2}{\hat s + 2m_-^2} L_1 \right\}
\eeq
\item[(ii)] \underline{$gg \to \squ \bar{\squ}$}: \\
\beq
\hat\sigma_{LO} = N_F \frac{\pi\alpha_s^2}{\hat s} \left\{
\beta_{\squ} \frac{5\hat s + 62 m_{\squ}^2}{24\hat s} -
\frac{m_{\squ}^2}{3\hat s} \frac{4\hat s + m_{\squ}^2}{\hat s} \log \left(
\frac{1+\beta_{\squ}}{1-\beta_{\squ}} \right) \right\}
\eeq
\underline{$q\bar q' \to \squ \bar{\squ}'$}: \\
\bea
\hat \sigma_{LO} & = & \frac{4\pi\alpha_s^2}{9\hat s} \left\{
\frac{\hat s + 2m_-^2}{\hat s} L_1 - \frac{m_{\gl}^2 + 2m_-^2}{m_{\gl}^2+m_-^2}
\beta_{\squ} \right. \\
& + & \left. \delta_{qq'} \left[ \frac{\beta_{\squ}}{3} \left(
\frac{1+2m_-^2}{\hat s} + N_F \frac{\hat s - 4m_{\squ}^2}{\hat s} \right)
+ 2 \frac{m_{\gl}^2\hat s + m_-^4}{\hat s^2} L_1 \right] \right\} \non
\eea
\item[(iii)] \underline{$gg \to \gl \gl$}: \\
\beq
\hat\sigma_{LO} = \frac{3\pi\alpha_s^2}{4\hat s} \left\{ 3\left(1-2
\frac{m_{\gl}^2}{\hat s} \right)^2 \log \left(
\frac{1+\beta_{\gl}}{1-\beta_{\gl}} \right) -
\frac{12\hat s+17 m_{\gl}^2}{\hat s} \beta_{\gl} \right\}
\eeq
\underline{$q\bar q \to \gl \gl$}: \\
\bea
\hat \sigma_{LO} & = & \frac{\pi\alpha_s^2}{\hat s} \left\{ \left(\frac{20}{27}
+ \frac{16 m_{\gl}^2}{9\hat s} - \frac{8 m_-^2}{3\hat s}
+\frac{32m_-^4}{27(m_{\gl}^2 \hat s + m_-^4)} \right) \beta_{\gl} \right. \\
& & \left. \hspace{1cm} + \left(\frac{64m_-^2}{27\hat s}
+\frac{8m_{\gl}^2}{27(\hat s-2m_-^2)}
-\frac{8}{3} \frac{m_{\gl}^2\hat s + m_-^4}{3\hat s^2} \right) L_2 \right\}\non
\eea
\item[(iv)] \underline{$gq \to \gl \squ$}: \\
\bea
\hat\sigma_{LO} & = & \frac{\pi\alpha_s^2}{\hat s}\left\{
-\frac{\lambda}{9\hat s^2} (7\hat s + 32 m_-^2)
+ \frac{2m_-^2}{9} \frac{4\hat s - m_{\squ}^2 -m_-^2}{\hat s^2} L_3
\right. \\
& & \left. \hspace{4cm}
+\frac{\hat s^2 + 2m_-^2\hat s - 2m_-^2 m_{\squ}^2}{\hat s^2} L_4 \right\} \non
\eea
\end{description}
with
\beq
\begin{array}{rclrcl}
L_1 & = & \displaystyle
\log\left(\frac{s+2m_-^2+\hat s\beta_{\squ}}{s+2m_-^2-\hat s\beta_{\squ}}
\right) \hspace{1cm}
& L_2 & = & \displaystyle \log\left(
\frac{s-2m_-^2+\hat s\beta_{\gl}}{s-2m_-^2-\hat s\beta_{\gl}} \right) \\ \\
L_3 & = & \displaystyle \log\left(\frac{s-m_-^2+\lambda}{s-m_-^2-\lambda}\right)
& L_4 & = & \displaystyle
\log\left(\frac{s+m_-^2+\lambda}{s+m_-^2-\lambda} \right) \\ \\
\beta_{\squ} & = & \displaystyle \sqrt{1 -\frac{4m_{\squ}^2}{\hat s}}
& \beta_{\gl} & = & \displaystyle \sqrt{1 -\frac{4m_{\gl}^2}{\hat s}} \\ \\
m_-^2 & = & m_{\gl}^2 - m_{\squ}^2
& \lambda & = & \displaystyle \sqrt{ (\hat s - m_{\gl}^2 - m_{\squ}^2)^2 -
4m_{\gl}^2 m_{\squ}^2}
\end{array}
\eeq
The LO hadronic cross sections can be obtained by convoluting the partonic ones
with the
corresponding parton densities. At the Tevatron $\squ \bar{\squ}$ production is
the dominant process for squark masses lighter than the gluino mass, while in
the reverse case gluino pair production dominates. The associated $\squ\gl$
production process always provides a sizeable fraction of SUSY
particle production. Beyond the excluded mass range the $q\bar q$ intial state
dominates over the $gg$ initial state in $\squ \bar{\squ}$ and $\gl \gl$ pair
production. At the LHC the situation is reversed: for smaller squark masses,
$\squ \squ$ and $\squ \gl$ production are dominant, while for larger squark
masses SUSY particle production is dominated by $\gl \gl$ and $\squ \gl$
production. Moreover, the $gg$ initial state clearly dominates gluino pair
production. The LO scale dependence leads to an estimate of $\sim 50\%$
uncertainty for
the production cross sections. Since the reconstruction of squarks and gluinos
will be very difficult due to the escaping LSP, an important possibility to
measure the squark and gluino masses is provided by the values of the total
cross sections. Thus a complete NLO calculation is needed in order to reduce
the theoretical errors to a reliable level \cite{sqgl}.

\section{SUSY QCD Corrections}
The evaluation of the SUSY QCD corrections will be exemplified for
$\squ \bar{\squ}$ production. These corrections split into two pieces, the
virtual
ones, generated by virtual particle exchanges, and the real ones,
which originate from gluon radiation and the corresponding crossed processes
with three-particle final states.

\subsection{Virtual corrections}
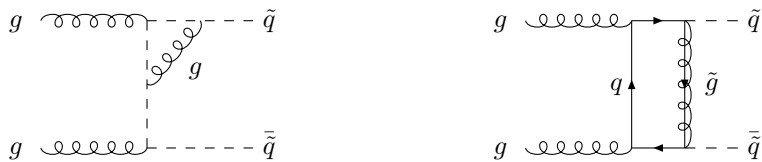
\begin{figure}[hbt]
\vspace*{-0.6cm}
\begin{center}
\SetScale{0.8}
\begin{picture}(120,60)(20,20)

\Gluon(0,20)(50,20){-3}{5}
\Gluon(0,80)(50,80){3}{5}
\Gluon(50,50)(75,80){-3}{4}
\DashLine(100,20)(50,20){5}
\DashLine(50,80)(100,80){5}
\DashLine(50,20)(50,80){5}
\put(-12,62){$g$}
\put(-12,14){$g$}
\put(56,44){$g$}
\put(84,14){$\sqb$}
\put(84,62){$\sq$}

\end{picture}
\begin{picture}(120,60)(-40,20)

\Gluon(0,20)(50,20){-3}{5}
\Gluon(0,80)(50,80){-3}{5}
\Gluon(75,20)(75,80){-3}{5}
\ArrowLine(75,80)(75,20)
\ArrowLine(75,20)(50,20)
\ArrowLine(50,20)(50,80)
\ArrowLine(50,80)(75,80)
\DashLine(100,20)(75,20){5}
\DashLine(75,80)(100,80){5}
\put(-12,62){$g$}
\put(-12,14){$g$}
\put(32,38){$q$}
\put(68,38){$\gl$}
\put(84,14){$\sqb$}
\put(84,62){$\sq$}

\end{picture}  \\
\caption[]{\label{fg:virt} \it Typical diagrams of the virtual corrections.}
\end{center}
\end{figure}
The one-loop virtual corrections are built up by gluon, gluino, quark and
squark exchange contributions [see Fig.~\ref{fg:virt}]. Their contraction with
the LO matrix elements provides the one-loop contributions to the physical
matrix elements. The evaluation of the virtual corrections has been performed
in dimensional regularization, leading to the extraction of
ultraviolet, infrared and collinear singularities as poles in
$\epsilon = (4-n)/2$. For the chiral $\gamma_5$ coupling we have used the naive
scheme, which is justified in the present analysis at the one-loop level
without anomalies.
We have explicitly checked that after summing all virtual corrections no
quadratic divergences are left over, in accordance with the general property
of supersymmetric theories. The renormalization has been performed by
identifying the squark and gluino masses with their pole masses, and defining
the strong
coupling in the $\overline{\rm MS}$ scheme, including five light flavors in the
corresponding $\beta$ function. The massive particles, i.e.\ squarks, gluinos
and top quarks, have been decoupled by subtracting their contribution at
vanishing momentum transfer \cite{decouple}. In dimensional regularization,
there is a mismatch between the gluonic degrees of freedom [d.o.f. = $n-2$] and
those of the gluino [d.o.f. = $2$], so that SUSY is explicitly broken. In
order to restore SUSY in the physical observables in the massless limit, an
additional finite counter-term is required for the renormalization of the novel
$\sq \gl \bar q$ vertex \cite{count}.

\subsection{Real corrections}
\begin{figure}[hbt]
\vspace*{-0.6cm}
\begin{center}
\SetScale{0.8}
\begin{picture}(120,70)(-20,10)

\Gluon(0,20)(50,50){-3}{5}
\Gluon(0,80)(50,50){3}{5}
\Gluon(25,65)(75,95){3}{5}
\DashLine(50,50)(100,80){5}
\DashLine(100,20)(50,50){5}
\put(-12,62){$g$}
\put(-12,14){$g$}
\put(64,74){$g$}
\put(84,14){$\sqb$}
\put(84,62){$\sq$}

\end{picture}
\begin{picture}(170,70)(-40,10)

\Gluon(0,20)(50,50){-3}{5}
\ArrowLine(0,80)(50,50)
\ArrowLine(50,50)(100,50)
\DashLine(100,50)(150,80){5}
\Gluon(100,50)(125,35){-3}{3}
\ArrowLine(100,50)(125,35)
\ArrowLine(125,35)(150,10)
\DashLine(125,35)(150,35){5}
\put(-12,62){$q$}
\put(-12,14){$g$}
\put(81,24){$\gl^*$}
\put(124,62){$\sq$}
\put(124,26){$\sqb$}
\put(124,6){$q$}

\end{picture}  \\
\caption[]{\label{fg:real} \it Typical diagrams of the real corrections.}
\end{center}
\end{figure}
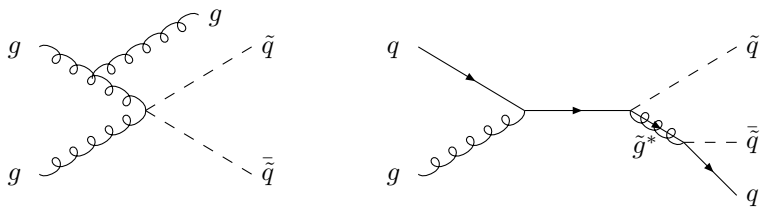
The real corrections originate from the radiation of a gluon in all possible
ways and from the crossed processes by interchanging the gluon of
the final state against a light quark in the initial state. The phase-space
integration of the final-state particles has been performed in $n=4-2\epsilon$
dimensions, leading to the extraction of infrared and collinear singularities
as poles in $\epsilon$. In order to isolate the singularities we have
introduced a cutoff $\Delta$ in the invariant mass of, say, the $\sqb g$ pair,
which separates soft and hard gluon radiation. After evaluating all
angular integrals and adding the virtual and real corrections, the infrared
singularities cancel. The left-over collinear singularities are universal and
are absorbed in the renormalization of the parton densities at NLO. We defined
the parton densities in the conventional $\overline{\rm MS}$ scheme including
five light flavors, i.e.\ the squark, gluino and top quark contributions are
not included in the mass factorization. Finally we end up with an ultraviolet,
infrared and collinear finite partonic cross section, which is independent
of the cutoff for $\Delta\to 0$.

However, there is an additional class of physical singularities, which have to
be regularized~\cite{sqgl}. In the second diagram of Fig.~\ref{fg:real} an
intermediate $\sq \gl^*$ state is produced, before the [off-shell] gluino splits
into a $q\sqb$ pair. If the gluino mass is larger than the common squark mass,
and the partonic c.m.\ energy is larger than the sum of the squark and gluino
masses, the intermediate gluino can be produced on the mass shell. Thus the
real corrections to $\sq \sqb$ production contain a contribution of $\sq \gl$
production. The residue of this part corresponds to $\sq \gl$ production with
the subsequent gluino decay $\gl \to \sqb q$, which is already contained
in the LO cross section
of $\sq \gl$ pair production, including all final-state cascade decays,
\beq
\frac{d\sigma_{res} (gq\to \gl \squ \to \sq \sqb q)}{d p_{\gl}^2}
= \hat \sigma(gq\to \gl \squ) BR(\gl \to \sqb q) \frac{m_{\gl}\Gamma_{\gl}/\pi}
{(p_{\gl}^2 - m_{\gl}^2)^2 + m_{\gl}^2 \Gamma_{\gl}^2}
\eeq
Thus this term has to be subtracted in order to derive a well-defined production
cross section for each individual final state. Analogous subtractions emerge
also in other reactions: if the gluino
mass is larger than the squark mass, the contributions from $\gl \to \sq \bar
q, \sqb q$ have to be subtracted, and in the reverse case the contributions of
squark decays into gluinos have to subtracted.

\begin{figure}[hbt]
\vspace*{-1.0cm}

\hspace*{0.5cm}
\epsfxsize=10cm \epsfbox{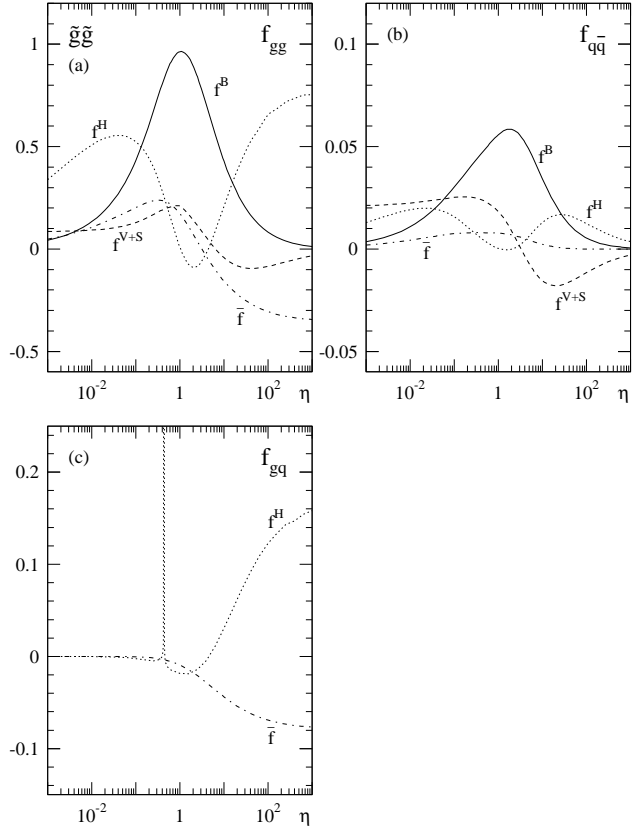}
\vspace*{-0.8cm}

\caption[]{\label{fg:scfc} \it Scaling functions of the gluino pair
production cross sections [$\eta = \hat s/(4m_{\gl}^2)-1$]. Top mass:
$m_t=175$ GeV.}
\end{figure}
\section{Results}
\subsection{Partonic cross sections}
The partonic scaling functions are defined as
\beq
\hat \sigma_{ij} = \frac{\alpha_s^2(Q^2)}{m^2} \left\{ f_{ij}^B
+ 4\pi \alpha_s (Q^2) \left[ f_{ij}^{V+S} + f_{ij}^H + \bar f_{ij} \log
\frac{Q^2}{m^2} \right] \right\}
\eeq
The parameter $m$ denotes the average mass of the produced massive particles.
The
scaling functions are shown for gluino pair production in Fig.~\ref{fg:scfc} as
a function
of $\eta = \hat s/(4m_{\gl}^2) - 1$ for the $gg, q\bar q$ and $gq$ initial
states. The singularity of $f_{gq}$ at $\hat s = (m_{\squ} + m_{\gl})^2$ is
a left-over singularity of the residue subtraction, which however is integrable
\cite{sqgl}.
The scaling functions can be evaluated analytically in the threshold region
$\beta \ll 1$ and the high energy limit $\hat s \gg m^2_{\squ}, m^2_{\gl}$.

In the threshold region the dominant contributions originate from Coulomb
singularities in the $f_{ij}^{V+S}$ functions due to Coulombic gluon exchange
between the final state particles and from large logarithmic enhancements in the
$f^H_{ij}$ and $\bar f_{ij}$ functions due to soft gluon radiation from the
initial states. The leading logarithms of soft gluon radiation can be resummed
\cite{resum}. The analytic results in the threshold region for e.g.\ $gg\to \sq
\sqb$ are given by
\beq
\begin{array}{rclrcl}
f^B_{gg} & = & \displaystyle \frac{7}{192} N_F \pi \beta & f_{gg}^{V+S} & = &
\displaystyle f_{gg}^B \frac{11}{336 \beta} \\ \\
f_{gg}^H & = & \displaystyle f_{gg}^B \left\{ \frac{3}{2\pi^2}
\log^2 (8\beta^2) - \frac{183}{28\pi^2} \log (8\beta^2) \right\} &
\bar f_{gg} & = & \displaystyle -f_{gg}^B \frac{3}{2\pi^2} \log (8\beta^2)
\end{array}
\eeq

In the high energy region the LO cross sections scale as $\hat \sigma_{LO}
\propto \alpha_s /\hat s$, while the NLO cross sections develop a high-energy
plateau, $\hat \sigma_{NLO} \propto \alpha_s^2/m^2$, originating from
$t$-channel gluon exchange between an initial quark/gluon line and the hard
process. Thus the scaling functions $f^H_{ij},\bar f_{ij}$ approach finite
values, which can be
calculated by using the factorization in $k_T$ of the exchanged gluon\cite{kt}.
For $\sq \sqb$ pair production the non-zero limits read as
\beq
\begin{array}{rclrcl}
f_{gg}^H & = & \displaystyle \frac{2159}{4320 \pi} \hspace*{1cm} & f_{gq}^H
& = & \displaystyle \frac{2159}{19440 \pi} \\ \\
\bar f_{gg} & = & \displaystyle -\frac{11}{72 \pi} & \bar f_{gq}
& = & \displaystyle -\frac{11}{324 \pi}
\end{array}
\eeq

\newpage

\subsection{Hadronic cross sections}
\begin{figure}[hbt]
\vspace*{-1.1cm}

\hspace*{0.5cm}
\epsfxsize=11cm \epsfbox{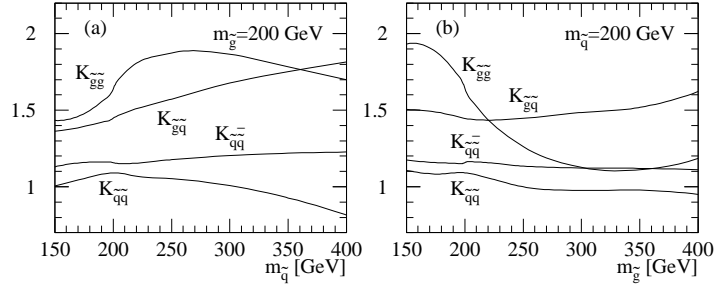}
\vspace*{-0.9cm}

\caption[]{\label{fg:kfactev} \it K factors of the different squark and gluino
production cross sections at the Tevatron. Parton density: GRV(94) with $Q=m$.
Top mass: $m_t=175$ GeV.}
\end{figure}
\begin{figure}[hbt]
\vspace*{-1.7cm}

\hspace*{0.5cm}
\epsfxsize=11cm \epsfbox{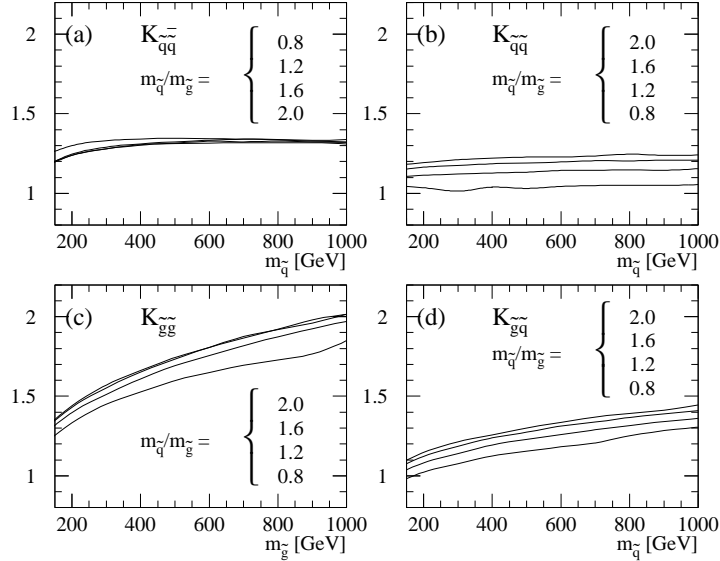}
\vspace*{-0.9cm}

\caption[]{\label{fg:kfaclhc} \it K factors of the different squark and gluino
production cross sections at the LHC. Parton density: GRV(94) with $Q=m$.
Top mass: $m_t=175$ GeV.}
\end{figure}
The hadronic cross sections can be obtained by convoluting the partonic ones
with the corresponding parton densities. We have performed the
numerical analysis for the Tevatron and the LHC. For the natural
renormalization/factorization scale choice $Q=m$, where $m$ denotes the
average mass of the final-state SUSY particles, the SUSY QCD corrections are
large and positive, increasing the total
cross sections by 10--90\% \cite{sqgl}. This is shown in
Figs.~\ref{fg:kfactev},\ref{fg:kfaclhc},
where the $K$ factors, defined as the ratios of the NLO and LO cross sections,
are presented as a function of the corresponding SUSY particle mass for the
Tevatron and the LHC.

\begin{figure}[t]
\vspace*{-7.3cm}

\hspace*{-1cm}
\epsfxsize=12cm \epsfbox{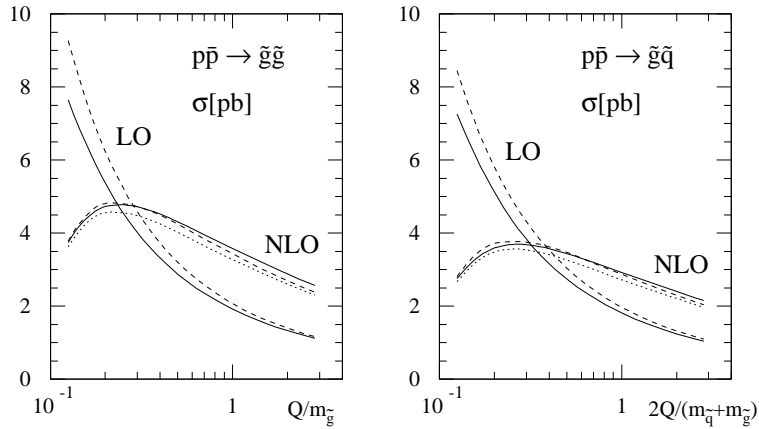}
\vspace*{-1.0cm}

\caption[]{\label{fg:scale} \it Scale and parton density dependence of the
total $\gl \gl$ and $\gl \sq$ production cross sections at the Tevatron in LO
and NLO. Parton densities: GRV(94) (solid), CTEQ3 (dashed) and MRS(A')
(dotted); mass parameters: $m_{\sq}=280$ GeV, $m_{\gl}=200$ GeV and $m_t=175$
GeV.}
\vspace*{-0.5cm}
\end{figure}
\begin{figure}[hbt]
\vspace*{-1cm}

\hspace*{-0cm}
\epsfxsize=9cm \epsfbox{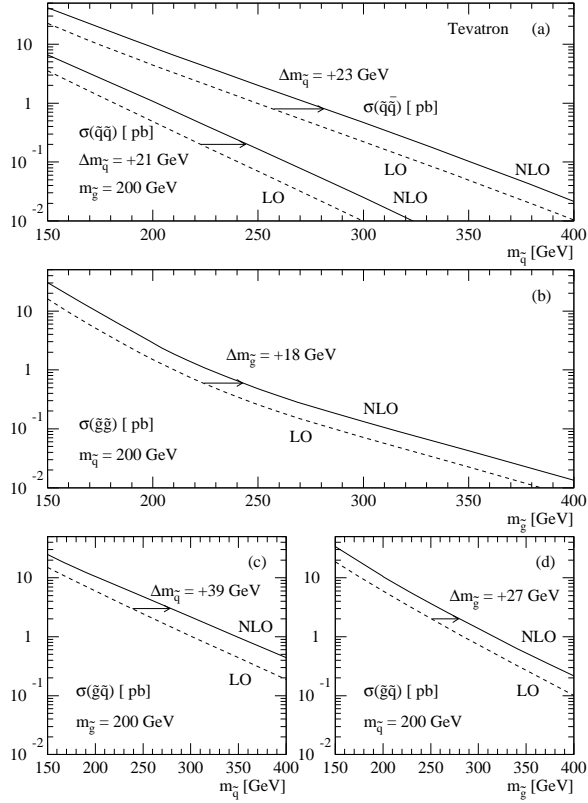}
\vspace*{-0.7cm}

\caption[]{\label{fg:masstev} \it The total cross sections for the Tevatron
[$\sqrt{S} = 1.8$ TeV]. NLO (solid): GRV(94) parton densities, with scale
$Q=m$; compared with LO (dashed): EHLQ parton densities, at the scale $Q =
\sqrt{s}$.}
\vspace*{-0.5cm}
\end{figure}
\begin{figure}[hbt]
\vspace*{-1cm}

\hspace*{-0cm}
\epsfxsize=9cm \epsfbox{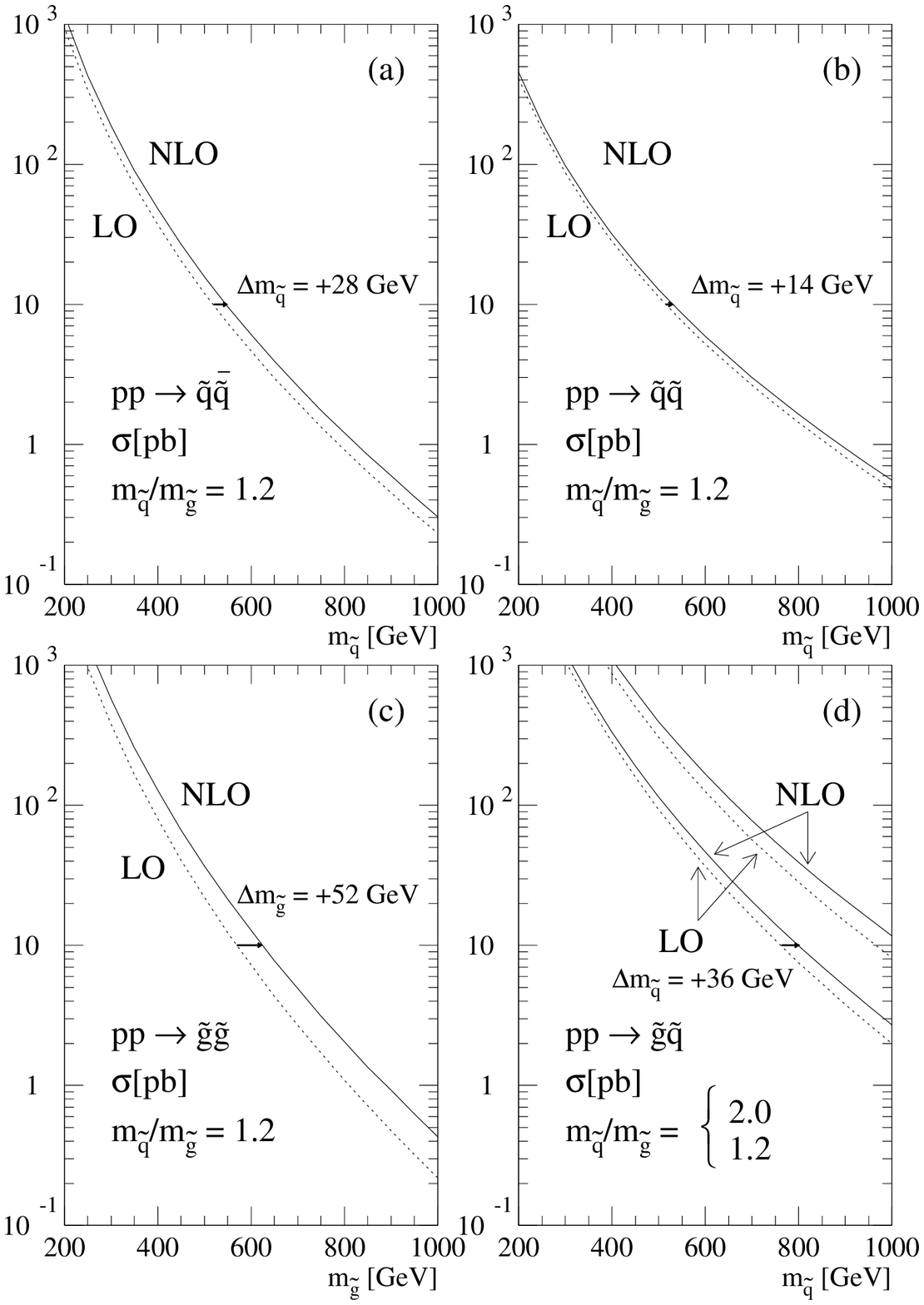}
\vspace*{-0.7cm}

\caption[]{\label{fg:masslhc} \it The total cross sections for the LHC
[$\sqrt{S} = 14$ TeV]. NLO (solid)
compared with LO (dashed). GRV(94) parton densities, with scale $Q = m$.}
\vspace*{-0.5cm}
\end{figure}

We have investigated the residual scale dependence in LO and NLO, which is
presented in Fig.~\ref{fg:scale}. Including
the NLO corrections reduces the LO scale dependence by a factor of 3--4 and
reaches a typical level of $\sim 15\%$, which serves as an estimate of the
remaining theoretical uncertainties \cite{sqgl}. Moreover, the dependence on
different sets of parton densities is rather weak and leads to an additional
uncertainty of $\sim 10\%$ \cite{sqgl}. In order to quantify the effect of the
NLO corrections on the
search for squarks and gluinos at hadron colliders, we have extracted the
SUSY particle masses corresponding to several fixed values of the production
cross sections. These masses are increased by 10--30 GeV at the Tevatron [see
Fig.~\ref{fg:masstev}] and
by 10--50 GeV at the LHC [see Fig.~\ref{fg:masslhc}], thus enhancing the
present bounds on the squark and gluino masses significantly \cite{sqgl}.

\begin{figure}[hbt]
\vspace*{-1.2cm}

\hspace*{0cm}
\epsfxsize=12cm \epsfbox{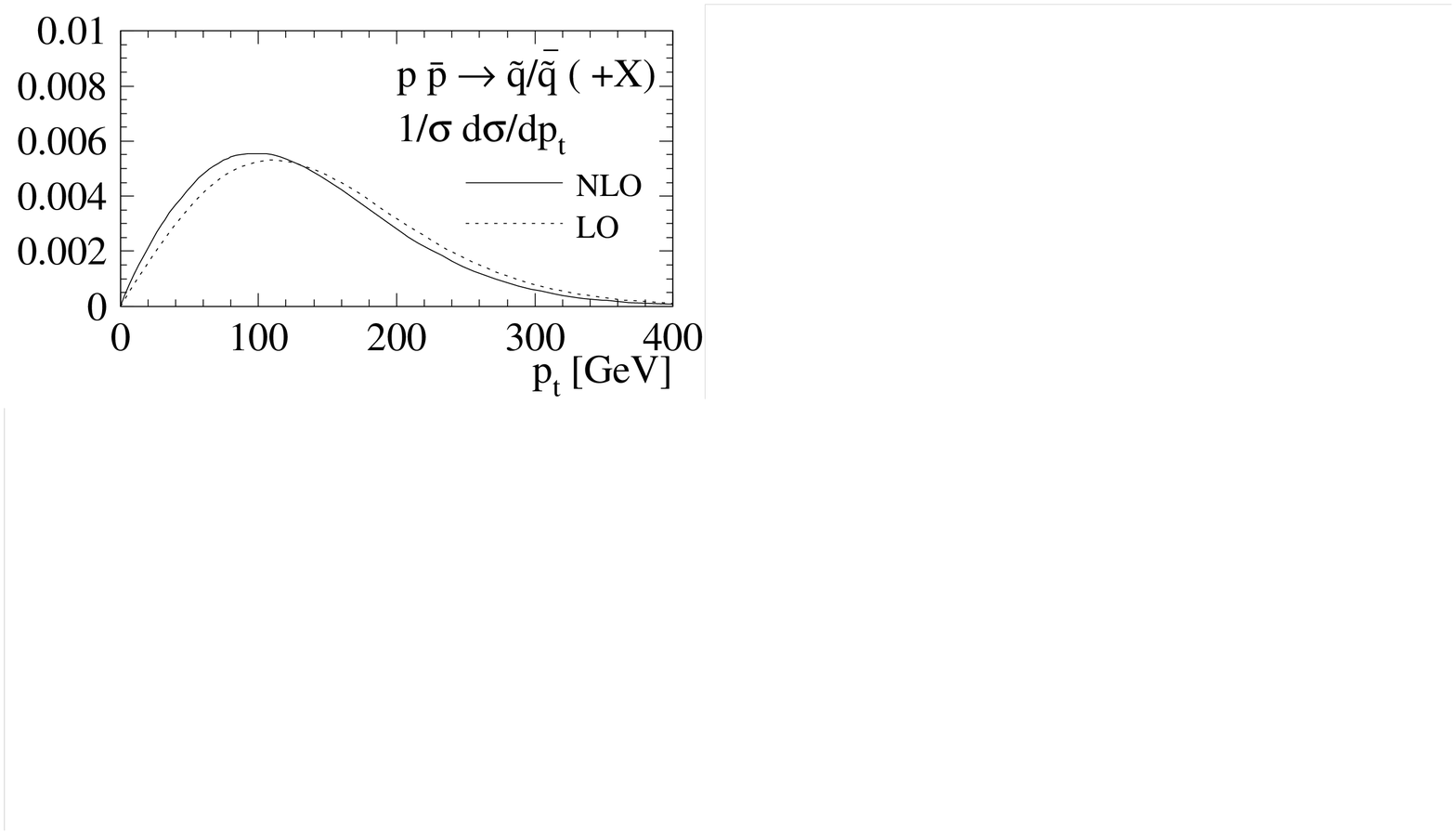}

\vspace*{-7.7cm}

\hspace*{6cm}
\epsfxsize=12cm \epsfbox{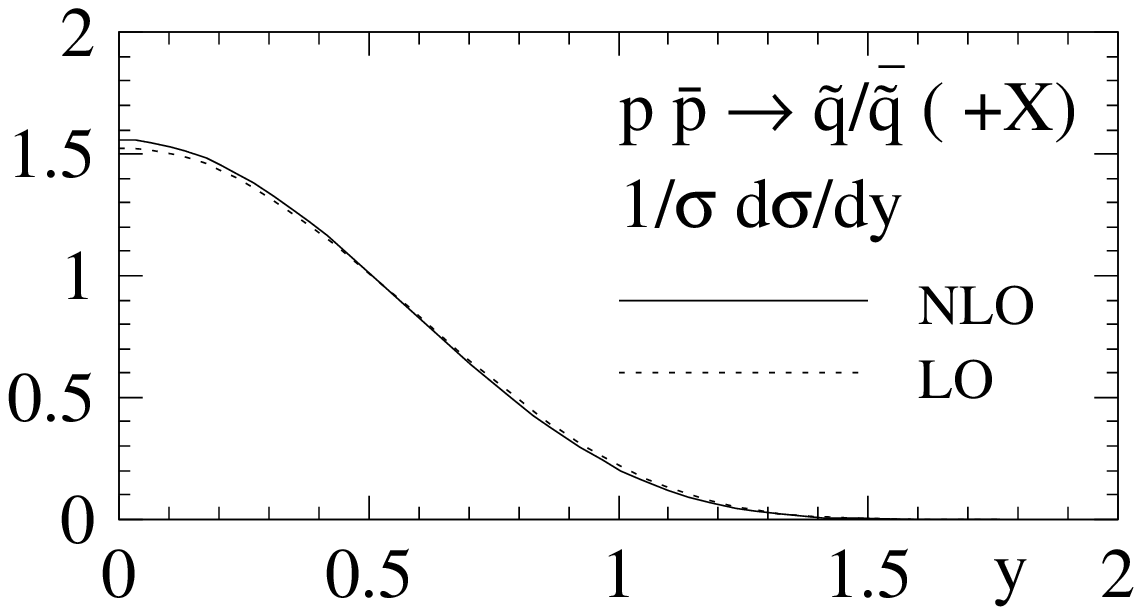}
\vspace*{-3.8cm}

\caption[]{\label{fg:pty} \it Normalized transverse-momentum and rapidity
distributions of $p\bar p\to \sq \sqb + X$ at the Tevatron in LO (dotted)
and NLO (solid). Parton densities:
GRV(94) with $Q=m$; mass parameters: $m_{\sq}=280$ GeV, $m_{\gl}=200$ GeV
and $m_t=175$ GeV.}
\vspace*{-0.5cm}
\end{figure}

Finally we have evaluated the QCD-corrected transverse-momentum and rapidity
distributions for all different processes. As can be inferred from
Fig.~\ref{fg:pty}, the modification of the normalized distributions in NLO
compared to LO is less than 10\% for the transverse-momentum
distributions and negligible for the rapidity distributions. Thus it
is a sufficient approximation to rescale the LO distributions uniformly by the
$K$ factors of the total cross sections \cite{sqgl}.

\section{Conclusions}
We have reviewed the theoretical status of the squark and
gluino production cross sections at present and future hadron colliders. The
evaluation of the full SUSY-QCD corrections has been described in detail.
They enhance the cross sections by about 10--90\% and are thus significant for
a reliable prediction. In spite of the large size of the NLO corrections, the
residual scale dependence reduced by a factor of 2.5--4, so that the theoretical
uncertainty decreased to $\sim 15\%$. The dependence on different parton
densities is $\sim 10\%$ and thus weak. The NLO corrections increase the
bounds on the squark and gluino masses, extracted from upper bounds on the total
cross sections, by about 15--35 GeV at the Tevatron and by 10--50 GeV at the
LHC. The shape of the differential distributions in the transverse momentum
$p_T$ and the rapidity $y$ are hardly affected by the NLO corrections, so that
a simple rescaling of the LO distributions by the $K$ factors of the total cross
sections provides a reasonable approximation within $\sim 10\%$. The final
results are encoded in the Fortran program PROSPINO
\cite{prospino}. \footnote{PROSPINO is available from
http://wwwcn.cern.ch/$\sim$mspira/ and
http://www-lorentz.leidenuniv.nl/wwreport/.}

\section*{Acknowledgements}
I would like to thank W.\ Beenakker, R.\ H\"opker and P.\ Zerwas for the
pleasant collaboration in the work presented here.

\end{document}